\documentclass[letterpaper, 10 pt, conference]{ieeeconf}  

\IEEEoverridecommandlockouts                              

\usepackage{graphicx}
\usepackage{wrapfig}
\graphicspath{ {images/} }

\title{Poster: Visual Content Privacy Leaks on Social Media Networks}

\author{Jasmine DeHart$^{1}$ and Christan Grant, PhD$^{2}$
\thanks{$^{1}$Jasmine DeHart is a Research Assistant with the Department of Computer Science,
        University of Oklahoma, Norman, OK, 73019 USA
        {\tt\small dehart.jasmine@ou.edu} supported by NSF Bridge to the Doctorate and OK-LSAMP}%
\thanks{$^{2}$Christan Grant is Faculty with the Department of Computer Science, University of Oklahoma,
        Norman, OK, 73019, USA
        {\tt\small cgrant@ou.edu}}%
}

\begin{document}

\maketitle
\thispagestyle{empty}
\pagestyle{empty}

\begin{abstract}

With the growth and accessibility of mobile devices and internet, the ease of posting and sharing content on social media networks (SMNs) has increased exponentially. 
Many users post images that contain ``privacy leaks'' regarding themselves or someone else. 
Privacy leaks include any instance in which a transfer of personal identifying visual content is shared on SMNs. 
Private visual content (images and videos) exposes intimate information that can be detrimental to your finances, personal life, and reputation. 
Private visual content can include baby faces, credit cards, social security cards, house keys and others. 
The Hawaii Emergency Agency example provides evidence that visual content privacy leaks can happen on an individual or organization level. 
We find that monitoring techniques are essential for the improvement of private life and the development of future techniques. 
More extensive and enduring techniques will allow typical users, organizations, and the government to have a positive social media footprint.

\end{abstract}

\section{Introduction}

\begin{figure}[h]
    \includegraphics[width=8.5cm, height=4cm]{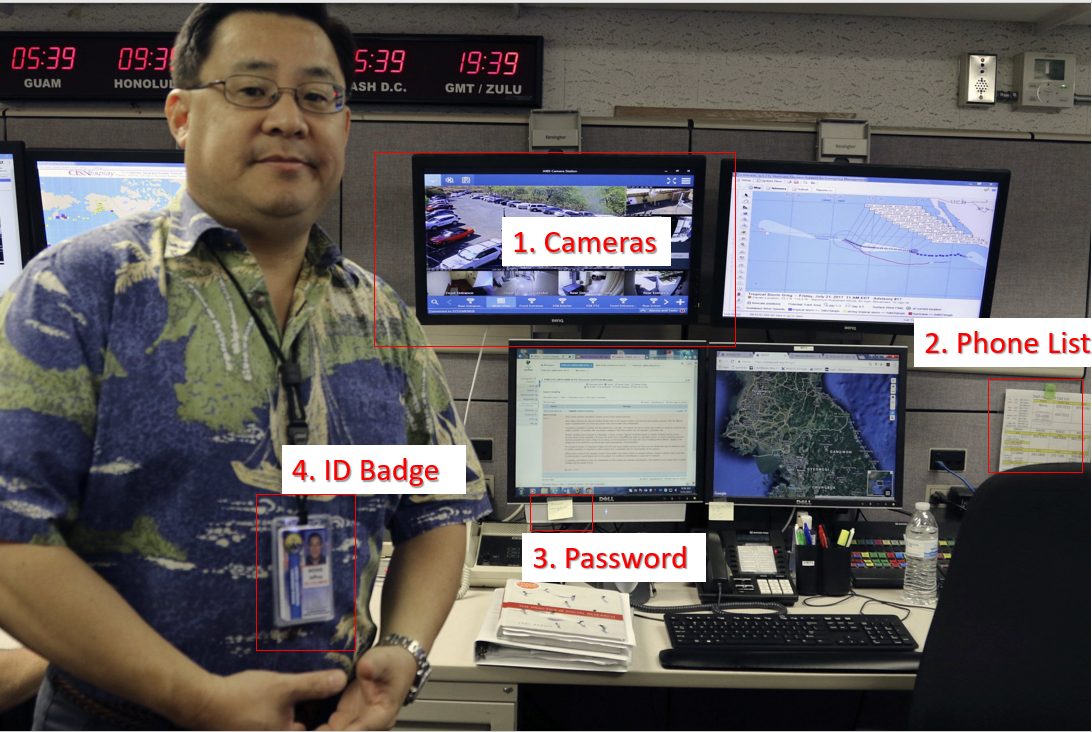}
    \caption{Hawaii Emergency Agency ~\cite{leswing2018password}}
    \label{fig:hawaiipic}
\end{figure}

According to Pew Research Center, 79 percent of Americans online use Facebook, 32 percent of Americans online use Instagram, and 24 percent of Americans online use Twitter ~\cite{greenwood2016social}.
Any content posted to social media networks (SMNs) can be lost to someone else even after removal of the content. 
Stolen visual content can then be used as a transport vector for other types of cyber-attacks or social engineering ~\cite{wilcox2016framework}. 
In one famous example, the Hawaii Emergency Management Agency released photo on Twitter that has raised several questions about the organization's cybersecurity policies and awareness ~\cite{leswing2018password}. 
In the Figure 1, it is noted that one of the monitors display a post-it note with a password (Figure~\ref{fig:hawaiipic}.3). 

This photo also exposes the location of the cameras (Figure~\ref{fig:hawaiipic}.1) which could lead to \textit{vandalism} and \textit{burglary}, a phone list (Figure~\ref{fig:hawaiipic}.2) which could lead to \textit{identity theft} and \textit{harassment}, and the badge of the employee (Figure~\ref{fig:hawaiipic}.4) which could lead to \textit{stalking/impersonation} ~\cite{hawaiitweet}.

\section{Goals}

In this work, we investigate (1) how pervasive social media-based privacy visual content leaks are and (2) what reasonable mitigation strategies can be developed to detect and minimize these leaks. We use deep learning techniques to identify ``private'' information. We propose and incorporate a privacy scoring metric to gauge a typical user's privacy leaks,which computes an individual user's probable exposure regarding their visual content leaks  ~\cite{grandison2017providing, liu2010framework}. On SMNs, content can be without any pre-screening or post-screening procedures, with this proposed procedure we are implementing screening mitigation techniques that will secure users' information. This endeavour will help us monitor visual content from SMNs for consumer privacy and protection. This research will subsequently protect everyday users from invasions of privacy, whether the action is accidental or intentionally made.

\section{Method}

\subsection{Categories}

Private visual content exposes intimate information that can be detrimental to our finances, personal life, and reputation. Private visual content can include baby faces, credit cards, phone numbers, social security cards, house keys and others. To further assess the sensitivity of  visual content, we have allocated these items into three categories.
\subsubsection{Location} An attacker can use this vector to find out where an individual lives and/or current location. The dangers of this include burglary, stalking, and kidnapping. Suggested methods to protect this information is to blur, black out, or to provide generic information. Examples include letters and phone numbers.
\subsubsection{Identity} An attacker can use this vector to exploit an individual's identity, even to the very intimate details. The dangers of this include identity theft, financial threat, burglary. Proposed methods used to protect this information is to black out, blur, or provide dummy information (Jane Doe, 00-00-0000...). Examples include birth certificate, driver’s license, and passports.
\subsubsection{Asset} An attacker can use this vector to exploit an individual's possessions and valuables. The dangers include financial threat, burglary, stalking, digital kidnapping, and online defamation. Proposed methods to protect this information are to blur and use dummy information (1234 4596 0000). Examples include baby faces, debit cards and keys.

\subsection{Mitigation Techniques}

\begin{figure}[h]
    \includegraphics[width=1\columnwidth]{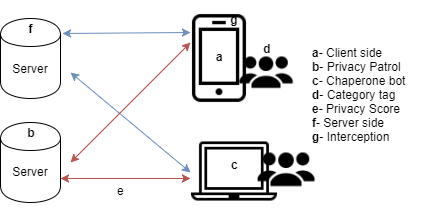}
    \caption{The location of each proposed mitigation technique.}
    \label{fig:monitoringtech}
\end{figure}

Given the various categories and examples in which visual content can be exploited, we are building a Visual Content Privacy Leak (VCPL) system that will further help us understand those risks and mitigate them. Our system collects data from SMNs (by crawling on social media and saving visual content), and then uses machine learning/deep learning object detection techniques to identify potential privacy leaks in the content collected. Using this system, we plan to build SMN clones that aim to score severity of privacy leaks on visual content per individual basis. To do so, we are are implementing the following seven mitigation techniques:

\noindent
\textbf{Technique 1} - Client side (Figure~\ref{fig:monitoringtech}a). A user can download a third-party application with various SMN applications on electronics to prevent the user from posting potential leaks. 
This third-party application will pre-screen  visual content (images, videos) before it can be posted on SMNs.

\noindent
\textbf{Technique 2} - Privacy Patrol (Figure~\ref{fig:monitoringtech}b). This SMN crawler will randomly look at a user's pages, screening for privacy leaks and alerting the user of various potential leaks.

\noindent
\textbf{Technique 3} - Chaperone bot (Figure~\ref{fig:monitoringtech}c). A user can add a chaperone bot as a friend on SMNs. 
The chaperone bot will give the user friendly suggestions based on type and frequency of privacy leaks on SMNs.

\noindent
\textbf{Technique 4} - Category Tagging (Figure~\ref{fig:monitoringtech}d). A user can select the category that the visual content belongs to before being uploaded to SMNs. Once tagged, an automated system will check for content compliance with tag. If it does not fit the category, the user is notified of new category tag options.

\noindent
\textbf{Technique 5} - Privacy Score (Figure~\ref{fig:monitoringtech}e). The user will be monitored based a privacy score. 
The bot will monitor the user's content after posting. In this case, a person who has a higher privacy score will be monitored more closely than some one who has a lower score. 

\noindent
\textbf{Technique 6} - Server side (Figure~\ref{fig:monitoringtech}f). The SMN will screen visual content before uploading to platform. 
We suggest collaboration with SMNs to provide enforcement of user compliance and techniques.

\noindent
\textbf{Technique 7} - Interception (Figure~\ref{fig:monitoringtech}g). With the SMN applications, users will agree to let the SMN intercept the camera and gallery to flag and block content that should not be selected for posting.

\subsection{Redaction Techniques}
\begin{figure}[h]
    \includegraphics[width=1\columnwidth]{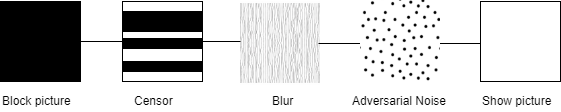}
    \caption{Spectrum of visual content redaction techniques}
    \label{fig:redaction-spectrum}
\end{figure}

Users may want to share images but hide all or parts of their images.
Web crawling systems may be collecting baby faces, credit card information for nefarious reasons.
We propose a spectrum of techniques to obfuscate images from users or machines (Figure~\ref{fig:redaction-spectrum}).

Once a visual content privacy leak is detected, we can handle it in various ways. 
The first option is to block the picture. This will remove the content and/or the user's affiliation with the content from SMNs. 
The second option is use a censor. 
Censoring is essentially removing a person or object from a visual content and it will insert a blank space where the object once existed. 
The third option is to blur content. 
Blurring the content will allow the user to have some control over what is being seen without causing too much distortion. The surrounding objects will still remain visible and the leaked object will be less visible but not removing them. 
The fourth option is to use adversarial noise ~\cite{goodfellow2014explaining}. 
We believe that adversarial noise will be important feature added to visual content to help protect SMNs user from computer attacks. 
By adding a few pixels, we could (1)  impede their ability to learn anything from the visual content even if it is in their possession, and (2) still allow the images to be visible to humans. 

\section{Conclusion}

As SMNs continue to grow in popularity, they become a powerhouse for privacy leakage due to the change in social culture, development of features, and audience. 
This research will impact everyday users and non-users of SMNs by providing a mechanism to identify sensitive information found in visual content posted on SMN. With the improvements in identifying and understanding privacy leaks on SMNs, we can lower the amount of malicious, financial and personal, attacks made on these platforms. 

\addtolength{\textheight}{-12cm}   







\bibliographystyle{ieeetr}
\bibliography{root}

\end{document}